\documentstyle[prd,aps]{revtex}
\begin{document}
\draft
\twocolumn[\hsize\textwidth\columnwidth\hsize\csname
@twocolumnfalse\endcsname
\preprint{FERMILAB--Pub--96/078-A,
SU-ITP-96-27,
{}~hep-ph/9606416}
\title{Preheating,  Supersymmetry Breaking
 and Baryogenesis}
\author{Greg.W. Anderson,$^{(1)}$ Andrei Linde,$^{(2)}$ and Antonio
Riotto$^{(1)}$}
\address{$^{(1)}${\it  Fermilab
National Accelerator Laboratory, Batavia, Illinois~~60510}}
\address{$^{(2)}${\it Department of Physics, Stanford University,
Stanford, California~~94305-4060}}
\date{June 18, 1996}
\maketitle
\begin{abstract}
Fluctuations of scalar fields produced at the stage of preheating after
inflation are so large that they can break supersymmetry much stronger than
inflation itself.   These fluctuations may lead to symmetry restoration along
flat directions of the effective potential even in the theories where the usual
high temperature corrections   are exponentially suppressed.  Our
results show that nonthermal phase transitions after preheating  may play a
crucial role in the  generation of the primordial baryon asymmetry by  the
Affleck-Dine mechanism. In particular, the baryon asymmetry may be generated at
the very early stage of the evolution of the Universe, at the preheating era,
and
not when the Hubble parameter becomes of order the gravitino mass.
\end{abstract}
\pacs{PACS: 98.80.Cq    \hskip 1 cm FERMILAB--Pub--96/078-A \hskip 1 cm
SU-ITP-96-27 \hskip 1 cm
 hep-ph/9606416}
\vskip2pc]

In the low-energy minimal supersymmetric standard model (MSSM) there exist a
large number   of $D$-flat directions along which squark, slepton and Higgs
fields may get expectation values. We collectively denote them by $\varphi$.
In
flat space at zero temperature
exact supersymmetry guarantees that the effective potential  $V(\varphi)$ along
these $D$-flat directions vanishes to all orders in perturbation theory
(besides the possible presence of nonrenormalizable terms in the
superpotential).
In the commonly studied supergravity scenario, supersymmetry breaking may take
place in isolated hidden sectors \cite{nilles} and then gets transferred to the
other sectors by gravity. The typical curvature of  $D$-flat directions
resulting from this mechanism   is
$M^2_\varphi \sim\frac{\left|F\right|^2}{M_{\rm P}^2}$,
where $F$ is the vacuum expectation value (VEV) of the $F$-term breaking
supersymmetry in the hidden sector. In order to generate soft masses
of order of $M_W$ in the matter sector, $F$ is to be of order of $(M_WM_{\rm
P})$
and sfermion and Higgs masses   along flat directions turn out to be in the TeV
range.

$D$-flat directions   are
  extremely important in the Affleck-Dine (AD) scenario for   baryogenesis
\cite{ad} if large expectation values along flat vacua are present during  the
early stages of the evolving Universe. When calculating high temperature
corrections to the effective potential at
large $\varphi$ one can often neglect the renormalizable interactions of the
field
$\varphi$ with other fields, because the fields directly interacting with
$\varphi$
become too heavy to be excited by thermal effects.
Therefore    thermal effects do not push the field
$\varphi$ towards $\varphi = 0$ in the early universe if initially it was large
enough.

For this reason, in the original version of the AD scenario \cite{ad,adlinde}
it was assumed that the initial value of the scalar field $\varphi$ was
different in different parts of the universe, as in the chaotic inflation
scenario.
However, recently it was pointed out that initial position of the field
$\varphi$
could be fixed by corrections to the effective potential which are proportional
to the Hubble parameter $H$. Indeed, in generic supergravity theories
moduli masses are of order of the Hubble parameter $H$  \cite{dine},
\cite{moduli}. One of the reasons is that the effective potential    provides a
nonzero energy density $V\sim\left|F\right|^2$  which
breaks supersymmetry. If the vacuum energy
dominates, the Hubble parameter is given by
$H^2=(V/3 M_{\rm P}^2)$ and therefore the curvature  along the $D$-flat
directions  becomes $M_\varphi^2=c H_{{\rm I}}^2$, where $c$ may be either
positive or negative. (Here we use  the reduce Planck mass $M_{\rm P} \sim
2\times
10^{18}$ GeV.) A similar  contribution  appears after inflation as well. For a
mechanism of stronger supersymmetry violation see \cite{Dvali}.

The most relevant part of the entire AD potential  in presence of
nonrenormalizable superpotentials
of the type $\delta W=(\lambda/n M^{n-3})\varphi^n$ somewhat schematically can
be represented as follows
\cite{dine}:
\begin{eqnarray}\label{AD}
V(\varphi)  &=&    {c_1 }\,m_{3/2}^2\, |\varphi|^2 - {c_2 } \:H^2|\varphi|^2  +
\frac{A\lambda H  \varphi^n  e^{i \,n\theta} + h.c.}{n\:M^{n-3}}\nonumber  \\
 &+& \lambda^2{|\varphi|^{2n-2}\over   M^{2n-6}} \ .
\end{eqnarray}
Here $c_i$ are some positive constants  which we  suppose to be $O(1)$,
$m_{3/2} = O(1)$ TeV is the gravitino mass, $M$ is some large mass scale such
as the GUT or Planck
mass, $n$ is some integer, which may take such values as $4$, $6$, etc., and
the $A$-term violates the baryon or the lepton number and has a definite
$CP$-violating phase relative to $\varphi$. This leads to symmetry breaking
with $ \varphi \sim (H M ^{n-3}/\lambda)^{1/n-2}$. For definiteness we will
take here $n = 4$ and $M = M_{\rm P}$.

The main idea of ref. \cite{dine} is that in the early universe  the field
$\varphi$ will be fixed at $ \varphi  \sim \sqrt {H M_{\rm P} /\lambda}$. When
$H$
drops down to $m_{3/2}$, the terms proportional to $H$ become smaller than the
first term in (\ref{AD}), and the field $\varphi$ begins oscillating near
$\varphi =
0$. If the field $\varphi$ can be associated with flat directions for
squark-slepton fields,  then the
oscillations of the field $\varphi$ near $\varphi = 0$ in the presence of the
A-term of eq. (\ref{AD}) may produce baryon
asymmetry of the universe.

The aim of this Letter is to show that there is a much stronger mechanism of
supersymmetry breaking in the early universe. This mechanism  may lead to
corrections to the quadratic terms in the effective potential of the AD field
which are much greater than $ H^2\varphi^2$ and to  $A$-terms much greater
than the one of eq. (\ref{AD}). As a result, under
certain conditions the Affleck-Dime mechanism of baryon asymmetry generation
may become quite different from the  version proposed in \cite{dine}.

The mechanism of   supersymmetry breaking which we are going to discuss is
related to particle production at the first stage of  reheating   after
inflation.
Indeed,  Kofman,  Linde  and  Starobinsky  have recently pointed out  that a
broad parametric resonance, which under certain conditions occurs  soon after
the end of inflation, may lead to a very rapid decay of the inflaton field
\cite{explosive}.  The inflaton energy is released in the form of inflaton
decay products, whose occupation number is extremely large. They have energies
much smaller than the temperature that would have been obtained by an
instantaneous conversion of the inflaton energy density
into radiation. Since it requires several scattering times for the low-energy
decay
products to form a thermal distribution, it is rather reasonable to
consider the period in which most of the energy density of the
Universe was in the form of the nonthermal   quanta produced by
inflaton decay as a separate cosmological era, dubbed as preheating to
distinguish it from the subsequent  stages  of particle decay and
thermalization which can be described by the techniques developed in
\cite{tec}.
Several aspects of the theory of explosive reheating have been studied in the
case of slow-roll inflation \cite{noneq} and first-order inflation
\cite{kolb} and it has been recently proposed that non-thermal production and
decay of Grand Unified Theory bosons at the intermediate stage of preheating
may generate the observed baryon asymmetry \cite{bario}.

One of the most important consequences of the stage of preheating is the
possibility of  nonthermal phase transitions with symmetry restoration
\cite{KLSSR}, \cite{tkachev}.
These phase transitions appear due to extremely strong quantum corrections
induced by particles produced at the stage of preheating.
What is crucial for our considerations is that
 parametric resonance is a phenomenon peculiar of particles obeying
Bose-Einstein statistics.   Parametric resonant decay into fermions is very
inefficient because of Pauli's exclusion principle. This means that during
the preheating period the Universe is  populated exclusively by a huge number
of
  soft bosons, and the  occupation numbers of bosons and fermions
belonging to the supermultiplet coupled  to the inflaton superfield are
completely unbalanced. Supersymmetry is then strongly broken during the
preheating era and large loop corrections may arise since the usual
cancellation between diagrams involving bosons and fermions within the same
supermultiplet
is no longer operative.  Therefore all results obtained in  \cite{KLSSR},
\cite{tkachev} apply to the modification of the effective potential along the
flat directions. As we will see,  the curvature $V^{\prime\prime}(\varphi)$
during the preheating era often is much larger than the effective mass
$H^2$ that $D$-flat directions acquire in
the inflationary stage.

Let us consider, {\it e.g}.,  chaotic inflation scenario \cite{linde}  where
the
inflaton field  $\phi$ couples to a complex  $\chi$-field,
\begin{equation}
\label{w}
V=\frac{M^2_\phi}{2}\:\phi^2+  g^2\:\phi^2\:|\chi^2| \ .
\end{equation}
We take the inflaton mass $M_\phi\sim 10^{13}$ GeV in order for for the density
perturbations generated during
the inflationary era to be consistent with COBE data.
Inflation occurs during the slow rolling of the scalar field $\phi$ from its
very large value. Then it oscillates with an initial amplitude $\phi_0\sim
M_{\rm P}$. Within   a dozen oscillations the initial energy
$\rho_\phi\sim M_\phi^2\:\phi_0^2$ is transferred through the interaction
$g^2\:\phi^2 \chi^2$ to  {\it bosonic} $\chi$-quanta
in the regime of parametric resonance \cite{explosive}.  At the end of the
broad parametric resonance  the field $\phi$ drops down   to $\phi_e \sim
10^{17}$ GeV. An exact number depends logarithmically on coupling constants; we
will use  $\phi_e \sim 10^{17}$ GeV $\sim 5\times 10^{-2} M_{\rm P}$ for our
estimates. After this stage,  the universe is   expected to be   filled up with
$\chi$-bosons with very large occupation numbers $n_k \sim g^{-2}$, with
relatively small energy per particle, $E_\chi \sim   \sqrt{gM_\phi \phi_e} \sim
0.2 \sqrt{gM_\phi M_{\rm P}}$. It is especially important that the amplitude of
field
fluctuations
produced at that stage is very large \cite{explosive}, \cite{KLSSR},
\begin{equation}\label{r1}
\langle \chi^2 \rangle \sim  5\times10^{-2} g^{-1} M_\phi M_{\rm P} \ .
\end{equation}
If, for example, the energy of the inflaton field after preheating were
instantly thermalized, we would obtain a  much  smaller value $\langle \chi^2
\rangle \sim 10^{-4}  M_\phi M_{\rm P}$. In realistic models thermalization
typically takes a lot of time, and the value of $\langle \chi^2 \rangle$ after
complete thermalization is many orders of magnitude smaller than
(\ref{r1}).

Note, that   large fluctuations  (\ref{r1}) occur only in the bosonic sector of
the
theory, thus breaking supersymmetry at the quantum level. Anomalously large
fluctuations of $\langle \chi^2 \rangle$ lead to specific nonthermal
phase transitions in the early universe   \cite{KLSSR}, \cite{tkachev}. It will
be interesting to study possible consequences of this effect for supersymmetric
theories with flat directions.

Let us make the natural assumption that
the field $\varphi$ labelling the $D$-flat direction  couples to the generic
supermultiplet $\chi$ by a renormalizable interaction of the form
${\cal L}_{{\rm int}}=h^2|\varphi|^2|\chi|^2$ coming from some $F$-term in the
potential.
Reheating gives an additional contribution to the effective mass along the
$\varphi$-direction :
\begin{equation}
\label{broad}
\Delta M^2_\varphi \sim  2 \:h^2 \langle \chi^2 \rangle \sim
  10^{-1}\:\frac{h^2}{g}\: M_\phi\:M_{\rm P}.
\end{equation}
 As a result, the curvature of
the effective potential becomes large and positive, and symmetry rapidly
restores for $10^{-1}\:\frac{h^2}{g}\: M_\phi\:M_{\rm P} > c_2\:H^2$. At the
end
of preheating in our model $H \sim  2\times10^{-2} M_\phi$. Thus, for $c_2 =
O(1)$ symmetry becomes
restored for $h^2 {\
\lower-1.2pt\vbox{\hbox{\rlap{$>$}\lower5pt\vbox{\hbox{$\sim$}}}}\ } 2\times
10^{-8}
g$. In other words, supersymmetry breaking due to preheating is much greater
than the supersymmetry breaking  proportional to $H$unless the coupling
constant $h^2$ is anomalously small. This leads to symmetry restoration in the
AD model soon after preheating.

Let us check the applicability of our results. First of all,
parametric resonance takes place for $g\phi > M_\phi$, where $\phi \sim
10^{-2} M_{\rm P}$ at the end of preheating. This implies that $g >
10^{-4}$. Secondly, even though self-interactions of the $\chi$-field do not
terminate the resonance effect since particles remain inside the resonance
shell, creation of quanta different from $\chi$  may remove the decay products
of the inflaton away from the resonance shell. This leads us to consider two
different possibilities. The resonance does not stop if  scatterings are
suppressed by kinematical reasons, {\it i.e.}  if the non-thermal plasma mass
of the final states is larger than the initial energy of the $\chi$'s. If the
final states are identified with the $\varphi$-quanta, this happens for $h>g$.
Otherwise, scatterings occur, but are slow enough  to not terminate the
resonance if the interaction rate $\Gamma\sim n_\chi\sigma$, where $\sigma$
denotes the scattering cross section,  is smaller than the typical frequency of
oscillations at the end of the preheating stage $\sim g\langle
\chi^2\rangle^{1/2}$. Identifying again the final states with the $\varphi$'s,
this condition translates into  $h^2  {\
\lower-1.2pt\vbox{\hbox{\rlap{$<$}\lower5pt\vbox{\hbox{$\sim$}}}}\ } 10^2 g$,
which is compatible with the condition for symmetry restoration discussed
above.

There is another necessary condition in order to  have broad resonance and a
rapid  production of $\chi$-particles:  the constant
contribution $h|\varphi|$ to effective mass  of the field $\chi$ induced by the
condensate $|\varphi_0|$ should be  smaller than the typical energy of the
decay products $E_\chi$. This translates into a bound
on the coupling $\lambda$ when taking into account the constraint on $h$
discussed previously. The exact bound depends upon the choice of the index $n$
and  the mass $M$. For $n = 4$ and $M = M_{\rm P}$ one has the condition $h^2 <
3
g\lambda$. This condition simultaneously guarantees that the energy density of
the AD field $\varphi$ is smaller than the inflaton energy density at the end
of the broad resonance. All these constraints are not very restrictive. For
example, one may
take $h = 2g$ to satisfy the condition $h > g$. Then one has the constraints $g
> 10^{-8}$ and $\lambda > g$.

Let us now describe the dynamics of the $\varphi$-field during  preheating.
Under the conditions described above,   the AD field $\varphi$ rapidly rolls
towards the origin $\varphi = 0$ and makes fast oscillations about it with an
initial amplitude given by $\varphi\sim |\varphi_0|$. The
frequency of the oscillations is
of order of $(\Delta M^2_\varphi)^{1/2}$ and is much higher than the Hubble
parameter at preheating $\sim 10^{-1}\:M_\phi$. The field is underdamped and
relaxes to the origin after a few oscillations.

The crucial observation  we would like to
make here is that baryon asymmetry can be efficiently produced during this fast
relaxation of the field to the origin. Our picture is similar to that of ref.
\cite{dine}, but there are considerable differences. According to \cite{dine},
the baryon asymmetry is produced at very late times when the Hubble parameter
$H$ becomes of order of the gravitino mass $m_{3/2}\sim$ TeV. At this epoch the
field becomes underdamped and the   $CP$-violating $A$-terms in the Lagrangian
are comparable to the baryon-conserving ones. The field feels a torque from the
$A$-terms and spirals from  the initial point
$\varphi=|\varphi_0|\:{\rm e}^{i\theta}$ inward in the harmonic potential,
producing a baryon asymmetry $n_B/n_\varphi={\cal O}(1)$.

In our case baryon number production occurs at very early stages, during the
preheating era. This is possible because the same nonthermal effects generating
a large curvature at the origin give rise also to sizable $CP$-violating terms
along the AD flat direction. Indeed, under very general assumptions, one can
expect the presence in the Lagrangian of nonrenormalizable $CP$-violating
terms of the type
$\alpha (\chi^2/M_{\rm P}^{m-2})\:\varphi^m+{\rm h.c.}$, where   $m>2$. During
the
preheating period large amplitudes of the $\chi$-field
give rise to the $CP$-violating term
\begin{equation}\label{A}
\alpha  \frac{\langle\chi^2\rangle}{m M_{\rm P}^{m-2}}\:\varphi^m  e^{i m
\theta_\chi
}\:\:+\:{\rm h.c.},
\end{equation}
For $m=3$   we get  a $CP$-violating term ${A\varphi^3 e^{i m \theta_\chi
}\over 3} +{\rm h.c.}$ with
$A\sim  10^{-1}\:g^{-1}\alpha\,M_\phi$.

Note that fluctuations $\langle\chi^2\rangle$ were absent during inflation;
they increase very sharply  (exponentially) during
preheating, and therefore they begin strongly affecting the position of the
scalar field only at the end of the broad parametric resonance, when they reach
their maximal value. Thus, to the first approximation one has the following
initial conditions for the field $\varphi$ at the end of the broad resonance:
$|\varphi_0| \sim \sqrt{HM_{\rm P}/\lambda} \sim \sqrt{M_\phi M_{\rm
P}/10\lambda}$,
$\dot\varphi_0 = 0$. As for the complex phase $\theta$, its initial value is
determined by the $H$-dependent A-term (\ref{AD}). In general, this phase is
quite different from the phase determined by eq. (5). As a result, the field
$\varphi$ will spiral down to $\varphi = 0$ acquiring baryon charge density
$n_B = 2|\varphi^2|\dot\theta$.

 After the end of the stage of broad parametric resonance the
value of $\langle\chi^2\rangle$ decreases approximately as $a^{-2} \sim
t^{-1}$. As a result, both $\Delta M_\varphi$ and $A$ after preheating decrease
as $t^{-1}$, i.e. in the same way as the Hubble constant $H$ in the radiation
dominated universe after preheating. Therefore the equation  describing the
relaxation of the $\varphi$-field toward the origin
\begin{equation}
\ddot{\varphi}+ 3\:H\:\dot{\varphi}+\Delta
M^2_\varphi\:\varphi+A\:(\varphi^*)^2=0
\end{equation}
takes the following form:
\begin{equation}\label{motion}
\ddot{\varphi}+ {3\over 2t} \dot{\varphi}+{1\over 2 gt} (h^2 M_{\rm P}\varphi
+
\alpha \varphi^{*2})=0 \ .
\end{equation}

Before going further, let us compare the magnitudes of the last two terms for
$|\varphi_0| \sim \sqrt{HM_{\rm P}/\lambda} \sim \sqrt{M_\phi M_{\rm
P}/10\lambda}$. One
can easily see that the $B$-violating term is greater than the mass term for
$\alpha > 10^3 h^2\sqrt\lambda$. If, e.g., one takes $h \sim 10^{-7}$, $\lambda
\sim 10^{-2}$, one finds that baryon number violation may be substantial even
if
the coefficient $\alpha$ in the A-term (\ref{A}) is extremely small, $\alpha
\sim 10^{-12}$.

It is convenient to introduce new variables $y = \varphi\sqrt {\lambda t
/M_{\rm P}}$ and $\tau = h\sqrt{2tM_{\rm P}/g}$. In these variables eq.
(\ref{motion}) simplifies:
\begin{equation}\label{motion2}
y''+ y+{\alpha\sqrt 2\over \tau\, h\, \sqrt {g\lambda}}\, y^{*2}=0 \ .
\end{equation}
In the new variables the motion of the field $y= |y|e^{i \theta}$ begins at
$|y_0| = 1$, at the initial moment $\tau_0 \sim  10^4 h/\sqrt g$. The condition
of vanishing initial velocity of the field $\varphi$ translates into the
condition $y'_0 = y_0/\tau_0$.   To calculate the
ratio  $n_B/s$, where  $n_B= 2|\varphi|^2\dot{\theta}$, and $s$ is the entropy
density, we will introduce a fictitious reheating temperature $T_R \sim
10^{-2} \sqrt {M_\phi M_{\rm P}} \sim 5\times 10^{13}$ GeV. This is the
temperature
which would be reached by our system if thermalization would occur
instantaneously after the end of the broad parametric resonance at $\phi_e \sim
10^{-1} M_{\rm P}$. Even though thermalization may occur much later, this
concept may
be quite useful because at the radiation dominated stage the energy density of
the universe decreases in such a way that at the moment when thermalization
actually occurs the resulting temperature $T$ will be equal to the  redshifted
value of the fictitious reheating temperature,  $T \sim T_R\sqrt{\tau_0/\tau}$.
 This leads to the following expression for the baryon number  $B = n_B/s$ soon
after preheating:
\begin{equation}\label{motion3}
B = {n_B\over s} \sim \frac{n_B}{10^2\,T_R^3} \sim  2\times10^{-2}\,  |y|^2\,
\theta' {h \over  \lambda \sqrt {g}} \ .
\end{equation}
We solved eq. (\ref{motion2}) numerically for various values of parameters and
calculated the ratio of $n_B$ to the entropy density $s$. We have found that
this ratio  oscillates and  approaches a constant at large times. For
${\alpha \over   h  \sqrt {g\lambda}} < 1$ and $h > 10^{-4} \sqrt g$ the
typical value of the baryon asymmetry is given by
\begin{equation}\label{motion4}
B = {n_B\over s}   \sim 10^{-2} {\alpha\over
g\lambda\sqrt \lambda}f(\tau_0)\ .
\end{equation}
Here $f(\tau_0)$ is a certain function of $\tau_0 \sim  10^4 h/\sqrt
g$\,:
 $f(1) \sim 1$, $f(10) \sim 0.1$, $f(100)  \sim 0.05$.

Validity of this result depends on details of the theory; typically it gives a
reliable estimate of the baryon asymmetry only for $B \ll 1$ (which is what we
need), but even in this case some care should be taken. For example, one can
show that for $h^2 < g\lambda$ the temperature $T$  remains much greater than
the time-dependent effective mass of the field $\varphi$ until this mass
approaches the constant value $O(m_{3/2})$. In this regime quarks acquire large
effective mass $O(T)$, and  the field $\varphi$ cannot decay and transfer its
baryon asymmetry to fermions until temperature drops to $O(m_{3/2})$. This may
lead to some corrections to eq. (\ref{motion4})  \cite{adlinde}. Also, the
simple scaling rules used in the derivation of eq. (\ref{motion4}) may break
if, e.g.,
at some intermediate stage the universe becomes dominated by nonrelativistic
particles.

A detailed investigation of baryon asymmetry production in realistic theories
including all of the effects mentioned above should become a subject of a
separate investigation. The main purpose of our paper was to show that
parametric resonance and nonthermal phase transitions found in
\cite{explosive,KLSSR} may lead to strong supersymmetry breaking in the early
universe and to considerable modifications in the theory of baryogenesis in
supersymmetric models. We have found that if the inflaton field $\phi$ couples
in a renormalizable way  to  $\chi$-bosons, which are weakly coupled to the AD
field $\varphi$, then the effect of parametric resonance    may induce  very
large masses  for particles corresponding to flat directions of the AD
potential.   The same effect may induce large  terms violating baryon
conservation (\ref{A}). As a result, one can obtain large baryon asymmetry even
in the models where all relevant coupling constants are extremely small.
\vskip 0.3 cm

We would like to thank D. Lyth, L.A. Kofman and E. Kolb for many useful
discussions. GA is supported by the DOE under contract DE-AC02-76CH03000; AR is
 supported by
the DOE and NASA under Grant NAG5--2788;
 AL\ is supported in part
by the  NSF grants
PHY-9219345 and AST-9529225.

\end{document}